\documentclass[%
reprint, 
amsmath,
amssymb,
aps,
pre,
]{revtex4-1}

\usepackage{graphicx}% Include figure files
\usepackage{dcolumn}% Align table columns on decimal point
\usepackage{bm}% bold math
\usepackage{float}
\usepackage{xcolor}
\usepackage[abs]{overpic}
\usepackage{mathtools}
\usepackage{subcaption}
\usepackage{graphicx}
\usepackage{caption}

\begin{document}

\title{The Effect of Obstacles in Multi-Site Protein Target Search with DNA Looping}

\author{Cayke Felipe$^{2}$}
\author{Jaeoh Shin$^{1,4}$}
\author{Yulia Loginova$^{5}$}
\author{Anatoly B. Kolomeisky$^{1,2,3,4}$}
\affiliation{$^1$Department of Chemistry, Rice University, Houston, Texas, 77005, USA}%
\affiliation{$^2$Department of Physics, Rice University, Houston, Texas, 77005, USA}%
\affiliation{$^{3}$Department of Chemical and Biomolecular Engineering, Rice University, Houston, Texas, 77005, USA}
\affiliation{$^4$Center for Theoretical Biological Physics, Rice University, Houston, Texas, 77005, USA}
\affiliation{$^5$Department of Chemistry, Moscow State University, Moscow, Russia}

\date{\today}

\begin{abstract}
Many fundamental biological processes are regulated by protein-DNA complexes called {\it synaptosomes}, which possess multiple interaction sites. Despite the critical importance of synaptosomes, the mechanisms of their formation remain not well understood. Because of the multi-site nature of participating proteins, it is widely believed that their search for specific sites on DNA involves the formation and breaking of DNA loops and  sliding in the looped configurations. In reality, DNA in live cells is densely covered by other biological molecules that might interfere with the formation of synaptosomes. In this work, we developed a theoretical approach to evaluate the role of obstacles in the target search of multi-site proteins when the formation of DNA loops and the sliding in looped configurations are possible. Our theoretical method is based on analysis of a discrete-state stochastic model that uses a master equations approach and extensive computer simulations. It is found that the obstacle slows down the search dynamics in the system when DNA loops are long-lived, but the effect is minimal for short-lived DNA loops. In addition, the relative positions of the target and the obstacle strongly influence the target search kinetics. Furthermore, the presence of the obstacle might increase the noise in the system. These observations are discussed using physical-chemical arguments. Our theoretical approach clarifies the molecular mechanisms of formation of protein-DNA complexes with multiple interactions sites.
\end{abstract}

\maketitle

\section{Introduction}
\label{sec-intro}

Various genetic modifications are critically important for genome maintenance of living systems and for ability of cellular organisms to adapt during the course of evolution \cite{alberts2014molecular,lodish2008molecular}. The examples of such processes include site-specific recombinations, gene regulation, genome rearrangements  and genome integration \cite{alberts2014molecular,grindley2006mechanisms,halford2004,mani2010triggers,bushman2005genome}. Large protein-DNA molecular assemblies known as {\it synaptic complexes} or {\it synaptosomes} typically control them \cite{alberts2014molecular,lodish2008molecular,grindley2006mechanisms,halford2004,cournac2013dna}. Although these complexes play important biological roles, it is still unclear how they  form so quickly and efficiently given the very complex nature of cellular environment \cite{grindley2006mechanisms}.  

The proteins in the synaptosomes can associate to DNA molecules at multiple sites. It is assumed that the overall process of formation of synaptosomes consist of several steps of sequential protein binding to the corresponding sites on DNA \cite{alberts2014molecular,lodish2008molecular,halford2004,cournac2013dna}. This suggests that the final steps of the synaptosome assembly take place when the protein is already bound to DNA at some locations, as shown in Fig. 1. Consequently, the search for the presently unoccupied specific sites should involve the formation of protein-facilitated DNA loops via non-specific protein-DNA interactions and protein sliding in the looped configurations. The appearance of such complex topological structures significantly complicates the understanding of the molecular mechanisms of the formation of synaptic complexes.

It is known that the single-site proteins find their specific sites on DNA via a combination of three-dimensional (3D) bulk diffusion and one-dimensional (1D) sliding of non-specifically bound proteins along the DNA chain \cite{halford2004,riggs1970lac,berg1976association,winter1981diffusion,von1989facilitated,coppey2004,lomholt2009,kolomeisky2013b,esadze2014positive,mirny2009,benichou2011,kolomeisky2011,sheinman2012}. This alternating mechanism can dramatically accelerate the search process, and it was intensively investigated in recent years \cite{mirny2009,kolomeisky2011,sheinman2012}. However, the appearance of DNA loops for multi-site proteins makes the search process significantly more complex. The theoretical methods used for investigating the target search of single-site proteins cannot be used for these systems. We recently introduced a theoretical framework for taking into account the appearance of DNA loops in the multi-site proteins target search \cite{kolomeisky2016,shin2019}. This theoretical approach uses a discrete-state stochastic description, and it combines analytical first-passage probabilities calculations with extensive Monte Carlo computer simulations. After identifying several dynamic search regimes, it was  shown that at some conditions the loop formation can lead to the accelerated target search \cite{kolomeisky2016,shin2019}. 

Large number of various protein molecules cover DNA in live cells \cite{alberts2014molecular,lodish2008molecular,marklund2013transcription,brackley2013}. This should influence the target search when the proteins slide along the DNA chain. The role of these biological obstacles has been investigated theoretically for the single-site proteins \cite{shvets2015role,shvets2016crowding,gomez2016facilitated}. These studies identified the mobility of obstacles and the strength of non-specific protein-DNA interactions as key features influencing the target search dynamics. However, the role of obstacles for multi-site proteins search has never been investigated.

In this paper, we present a theoretical study on the role of a static obstacle in the target search of multi-site proteins for specific sites on DNA, considering specifically the formation of DNA loops and sliding in the looped configurations. Using a combination of analytical and computer simulations methods, the multi-site proteins search dynamics is analyzed for general sets of conditions. Our theoretical calculations suggest that the lifetimes of DNA-looped configurations and relative positions of the target and the obstacle are the most important factors determining the search dynamics of multi-site proteins. In addition, adding the obstacle might increase the overall level of the noise in the system. Physical-chemical arguments to explain these observations are also presented.

\section{Theoretical Model}
\label{sec-model}

\subsection{Description}

A model for the multi-site protein target search with the obstacle is presented in Fig. 1. We will concentrate on the last stages of the formation of protein-DNA complex that involve DNA loops and sliding in the looped configurations. Let us consider is a single protein molecule with 2 binding sites, and one of them is already associated tightly to the end of the DNA chain with $L$ binding sites: see Fig. 1a. We set the size of each binding site to be 10 base pairs (bp) to reflect the typical size of the region of protein-DNA interactions.  One of these sites at the location $m$ ($1 \le m \le L$) is the target sequence for the second site on the protein molecule. In addition, there is an obstacle particle at the site $l \ne m$, that stays forever at the same location. In this study, we consider the static obstacles because, as was shown recently, they influence the search dynamics more than the mobile obstacles \cite{shvets2015role}.  Due to non-specific interactions, the protein molecule can non-specifically associate to DNA at any site, and this leads to the formation of DNA loops (Fig. 1a). In the looped conformation, the protein might slide along the DNA chain, but it cannot pass the obstacle.

It is important to note that since we are trying to understand the complex mechanisms of multi-site protein search using a minimal theoretical model, there are several approximations in our approach \cite{shin2019}. They include the assumption that the chain segments in the DNA loop can quickly relax to equilibrium, which is reasonable for not very long DNA chains. In addition, we assume that the location of non-specific bindings are not correlated. Furthermore, the possible appearance of DNA supercoiling and twists during the motion in looped configurations, as well as DNA sequence effects are also neglected. A more detailed discussions on these issues and on the validity of these approximations can be found in Ref. \cite{shin2019}. 

Now we can construct a discrete-state description of the model as presented in Fig. 1b. The overall system consisting of the single two-site protein bound to the end of DNA and the DNA chain can be viewed as $L+1$ discrete chemical states. The state $n$ ($1 \le n \le L$) describes the DNA looped state when the loop of size $n$ is formed and there are $L-n$ DNA sites that are not in the loop. For $n=0$, we have a state without loop, i.e., the protein is not associated to DNA via non-specific interactions (but recall that one site is always bound to the DNA end): see Fig. 1. From this unlooped conformation the protein molecule can associate to DNA at the site $n$ with a rate $k_{\text{on}}(n)$. The corresponding dissociation rate from the state $n$ is $k_{\text{off}}(n)$. The protein binding to DNA is associated with a binding energy $\epsilon$ (enthalpic contribution), where $\epsilon<0$ describe the attractions and $\epsilon>0$ correspond to repulsions. In the looped conformations, the protein can slide along the DNA with position-dependent rates: The transition $n \rightarrow n-1$ (reducing the loop size) is taking place with a rate $w_{n}$, while the transition $n \rightarrow n+1$ (increasing the loop size) is taking place with a rate $\mu_{n}$ (Fig. 1).

\begin{figure}[H]
 \centering
  % include first image
 \includegraphics[width=1.0\linewidth]{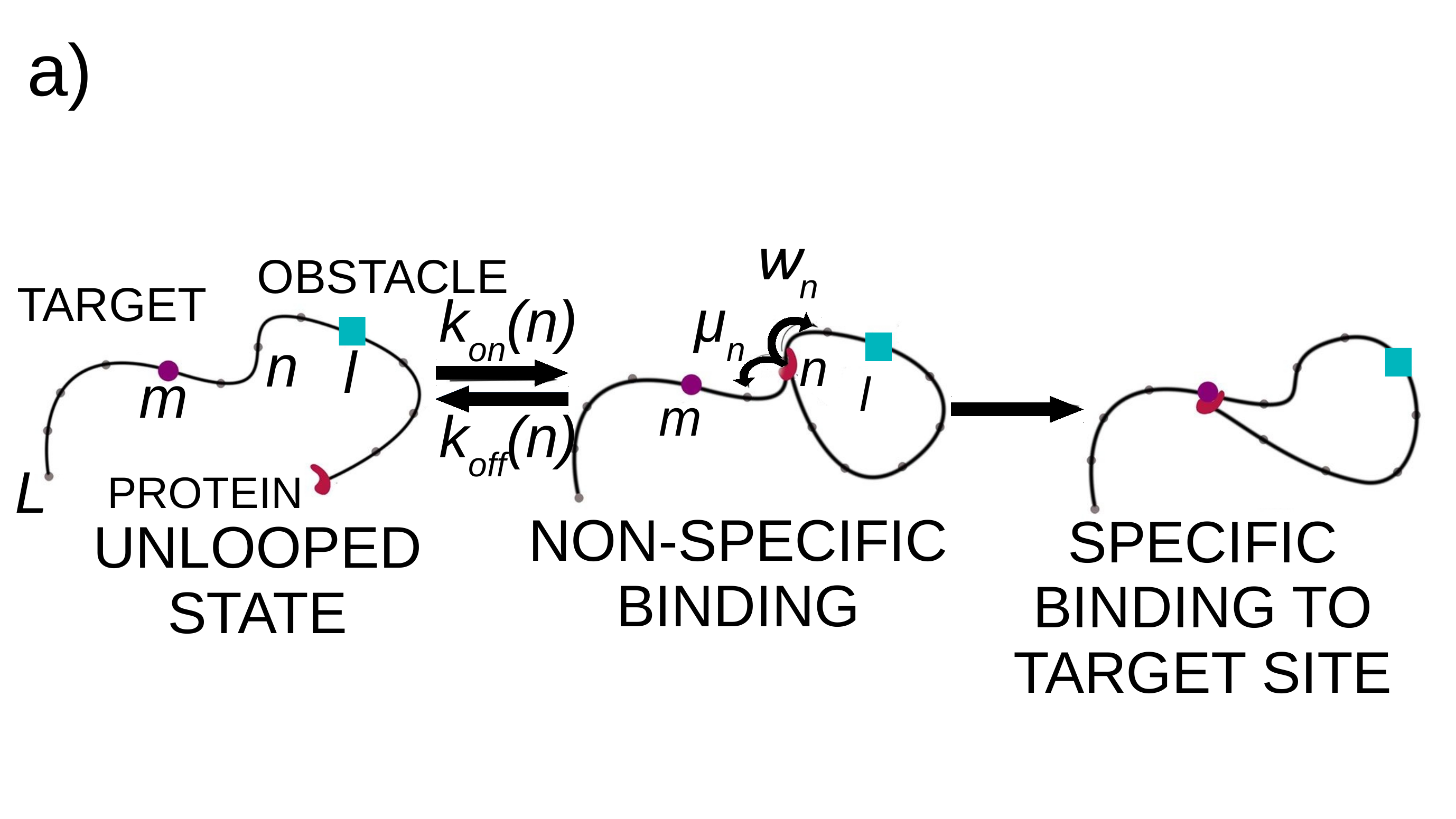} 
 \includegraphics[width=1.0\linewidth]{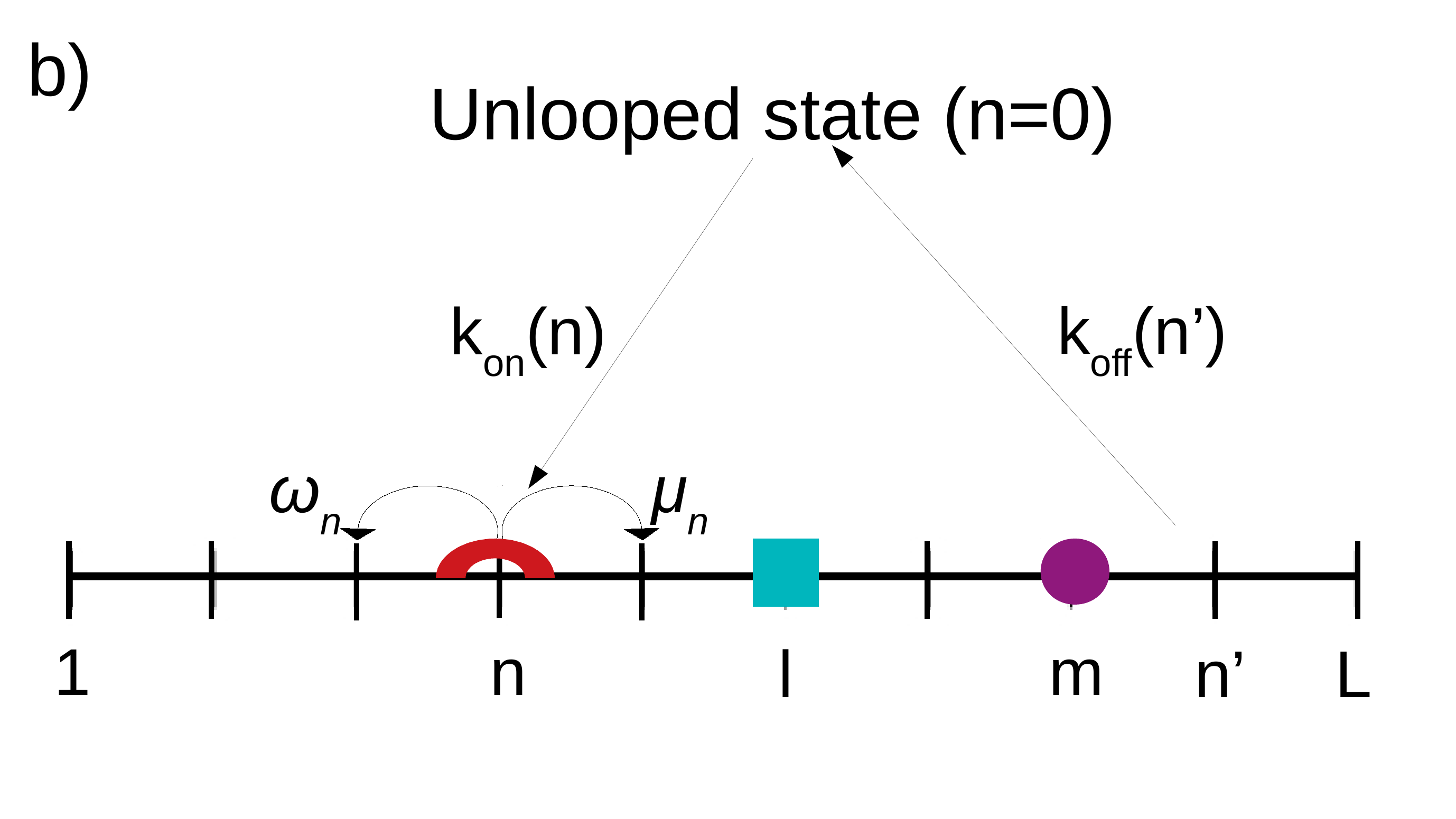}  
\caption{ a) Schematic view of two-site protein search for specific site on DNA when one site is already bound to DNA end. Red arc describes the proteins molecule. Purple circle  describes the target site. Cyan square describes  the obstacle particle.  Black arrows indicate allowed transitions. b) A corresponding discrete-state stochastic model of the search process.}
\label{fig:1}
\end{figure}

The formation of DNA loops and sliding in the loop configurations modify the free energy of the system, and this is the reason for position-dependent association/dissociation and sliding rates. If all these transitions are slower than the chain relaxation, the free energy of the system can be well approximated as \cite{shin2019}
\begin{equation}
    G(n)=G_{0}(n)+ \epsilon=\frac{A}{n} + \alpha\text{log}[n]+\epsilon,
    \label{eq:energy}
\end{equation}
where $G_{0}(n)$ is the free energy of the formation of the loop of size $n$ without enthalpic contribution. The first term in this expression corresponds to the bending energy of DNA, the second term describes  the entropy cost of creating the loop, and the last term is the enthalpic contribution due to the formation of non-specific protein-DNA bond. The coefficient $A$ is the bending stiffness of the DNA chain, and for a circular loop it is given by $A=2\pi^2 l_p$, where $l_p$ is the persistence length. Below, for specific calculations, we set the chain length as $L=300$ and the parameter $A=300$, which in  real units  correspond to $L=3000$ bps and $l_p=150$ bps. The coefficient $\alpha$ is related to a scaling exponent for the radius of gyration, and for an ideal Gaussian chain, it is equal to $\alpha=3/2$. The free energy profile for these parameters is presented in Fig. 2 (with $\epsilon=-2$ k$_{B}$T). We note that this free energy expression is one of the simplest approximations which neglects many details --- see Ref. \cite{shin2019} for more discussions on this issue.

\begin{figure}
\centering
\includegraphics[width=1.0 \columnwidth]{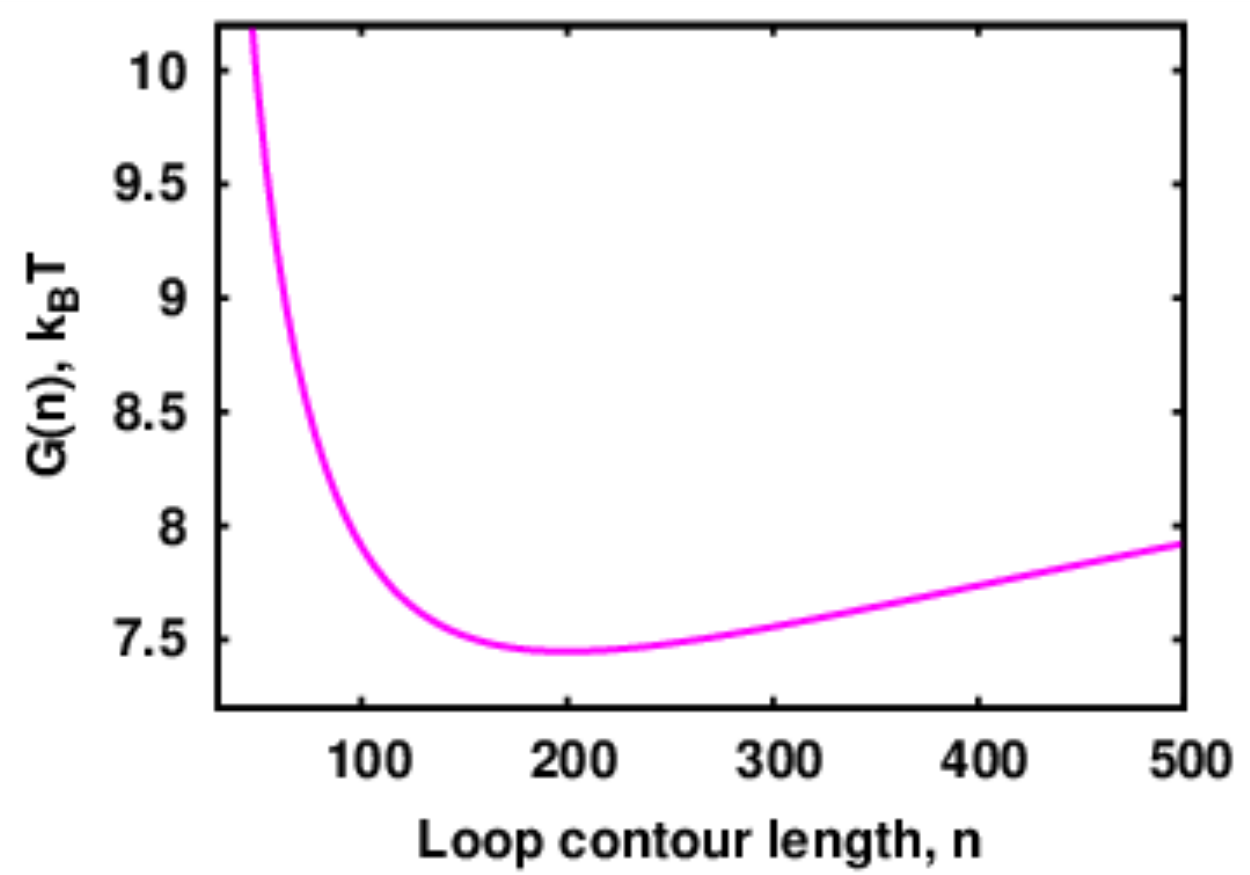}
\caption{Free-energy profile of the two-site protein target search as a function of the loop contour length $n$. The following parameters were used: $A=300$, $\alpha=3/2$ and $\epsilon=-2$ k$_{B}$T.}\label{fig-2}
\end{figure}

Knowing the free-energy profile of the system allow us to write explicit expressions  for the transition rates,
\begin{eqnarray}
    k_{\text{on}}(n) & = & k_{\text{on}}^{(0)}~\text{exp}[-\theta G_0(n)],\\
    k_{\text{off}}(n) & = & k_{\text{off}}^{(0)}~\text{exp}[(1-\theta) G_0(n)],\\
    \mu_n &= & \mu_0~\text{exp}[-\theta_t \Delta G(n+1)],\\
    w_n &= & \mu_0~\text{exp}[(1-\theta_t) \Delta G(n)].
\end{eqnarray}
Here, $k_{\text{on}}^{(0)}$, $k_{\text{off}}^{(0)}$, and $\mu_0$ are the binding, unbinding, and hopping rates, respectively, in the absence of the loop formation ($G(n)=0$), and $\Delta G(n) \equiv G(n)-G(n-1)$. Additionally, parameters $\theta$ and $\theta_t$ specify how the free energy change is distributed between the binding and unbinding transitions and for the sliding rates in different directions, respectively. For convenience, we set both of them to be $\theta=\theta_t=0.5$. It has been argued before that changing the values for these parameters does not change the qualitative behavior of the system \cite{shin2019}. In the following calculations,  we also take $\mu_0 \simeq 60$ s$^{-1}$ from the experimental measurements of the diffusion constant of p53 protein \cite{tafvizi2008tumor}, and both transition rates $k_{\text{on}}^{(0)}$ and $k_{\text{off}}^{(0)}$ are varied for a wide range of values.

 \subsection{Analytical Calculations in Limiting Cases}

It is not possible to explicitly analyze the discrete-state stochastic model for general sets of parameters. However, there are two limiting situations that allow for full analytical solutions, providing a valuable physical insights on the molecular mechanisms of the target search.

Let us start with the case of short-lived DNA loops. Then the protein molecule after binding to DNA does not have much time to slide and it quickly dissociates. Clearly, the obstacle does not affect much the dynamics of the target search. As the protein barely slides in this regime, it effectively does not encounter the obstacle at all times. This is a so-called ``no-sliding limit'' that was analytically fully investigated in Refs. \cite{shvets2016role,shin2019}. For the mean search time $T$, it was shown that   
\begin{equation}
T = \frac{1+\displaystyle\sum_{i\neq\{m,l\}}\frac{k_{\text{on}}(i)}{k_{\text{off}}(i)}}{k_{\text{on}}(m)}.
\end{equation}
This expression has a simple physical meaning. Because there are no correlations in binding/unbinding events, on average, the protein has to visit  every site on DNA (except the one occupied by the obstacle) before the target can be found. 

\begin{figure}[H]
\centering
\includegraphics[width=0.95\columnwidth]{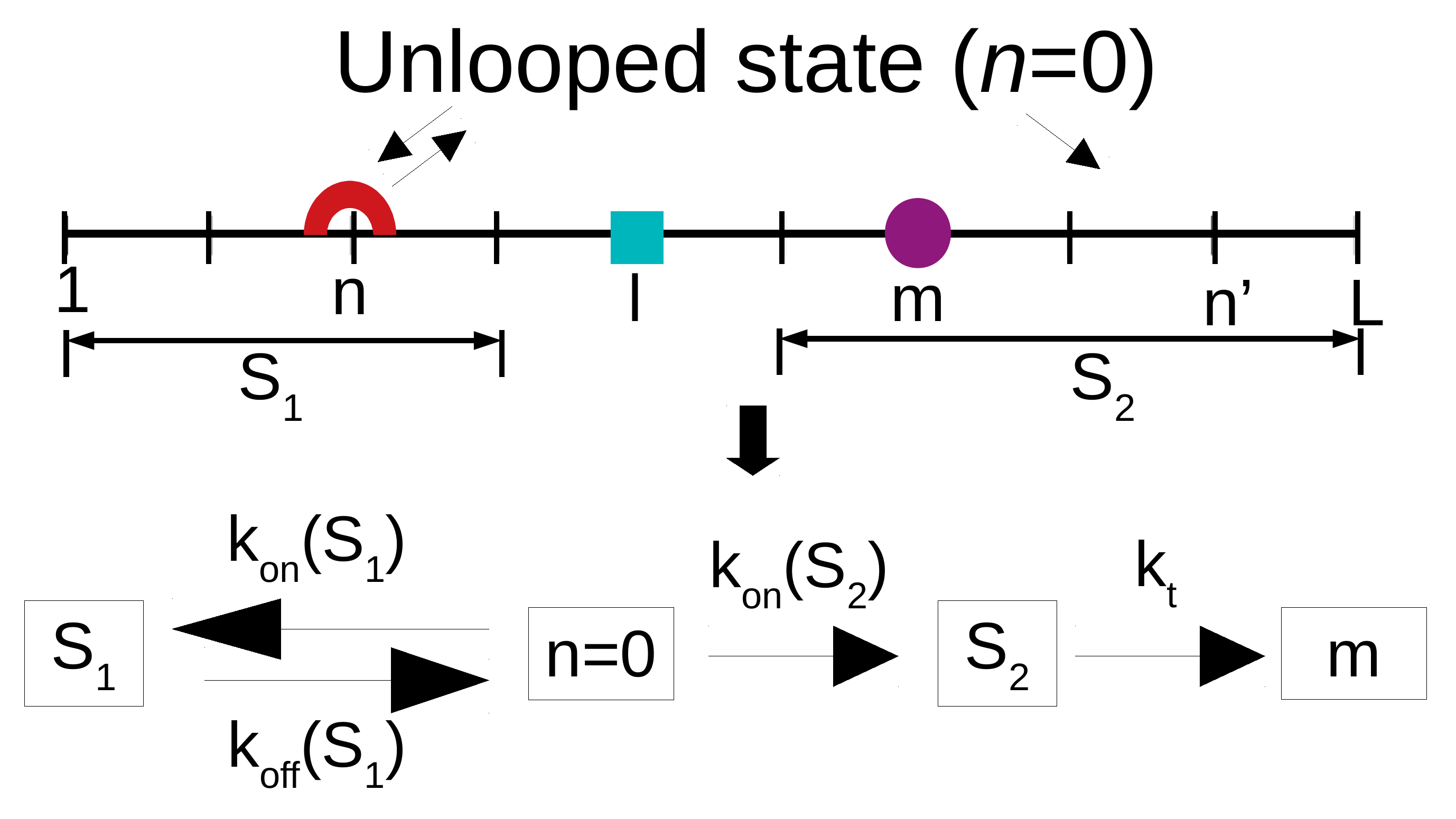}
\caption{A discrete-state stochastic model for the multi-site protein target search in the limit of long-lived DNA loops. The system is described as three distinct states, labeled as $S_{1}$, $n=0$ and $S_{2}$, respectively.}
\label{fig:diagram}
\end{figure}

Another limiting situation that can be analytically solved is the case with long-lived DNA loops. In this regime, once the protein binds it will stay on DNA for a long time. Then there are two possible situations. If the protein lands on the DNA segment where the obstacle does not prevent it from sliding to the target, it will eventually find it without dissociating. However, if the protein binds to the DNA segment where the obstacle is the barrier for reaching the target that cannot be passed the protein will spend a lot of time on this segment before dissociating. We can divide the DNA chain into two segments separated by the obstacle, as shown in Fig. 3. The region that does not include the target is labeled as $S_1$, and the other region is labeled as $S_2$: see Fig. \ref{fig:diagram}. We consider stationary state conditions  which means that as soon as the protein reaches the target site, the system starts immediately in the state $n=0$ \cite{shin2018surface}. From the steady-state flux $J$ to the target, the mean search time can be estimated as $T=1/J$. Let us define $P_0(t)$, $P_1(t)$ and $P_2(t)$ as the probabilities to be in the state $n=0$, in the state $S_1$ or in the state $S_2$, respectively, at time $t$. Then the dynamics in the system at all times is described by the following (forward) master equations,
\begin{eqnarray}
\frac{\partial P_0(t)}{\partial t} = & & -[k_{\text{on}}(S_1)+k_{\text{on}}(S_2)]P_0(t)+k_{\text{off}}(S_1)P_1(t)+k_t P_2(t);   \\
\frac{\partial P_1(t)}{\partial t}  = & & -k_{\text{off}}(S_1)P_1(t)+k_{\text{on}}(S_1)P_0(t);   \\
\frac{\partial P_2(t)}{\partial t} = &  &-k_{t}P_2(t)+k_{\text{on}}(S_2)P_0(t). 
\end{eqnarray}
In these expressions, $k_{\text{on}}(S_1)$, $k_{\text{on}}(S_2)$ and $k_{\text{off}}(S_1)$ are  transition rates to associate and dissociate from each region, which can be determined from the following expressions,
\begin{eqnarray}
k_{\text{on}}(S_1) & = & \sum\limits_{n \in S_1}k_{\text{on}}(n),\\  
k_{\text{on}}(S_2) & = & \sum\limits_{n \in S_2}k_{\text{on}}(n), \\
k_{\text{off}}(S_1) & = & \frac{\sum\limits_{n \in S_1}k_{\text{off}}(n)\text{exp}\left[ -G(n)\right]}{\sum\limits_{n \in S_1}\text{exp}\left[ -G(n)\right]}.
\end{eqnarray}
The physical meaning of these expressions is the following. The total rate to reach the segment $S_{1}$ or $S_{2}$ is the sum of all corresponding association rates to each segment.  The total dissociation rate out of the segment $S_{1}$ is the sum over all dissociation rates from all sites in the segment weighted by the probability to be found at each site. The parameter $k_t$ is the average rate to slide to the target at the site $m$ after reaching the region $S_{2}$. The explicit expression for this parameter can be obtained using the analysis developed in Ref. \cite{kolomeisky2013},
\begin{equation}
    k_{t}^{-1}  = \frac{1}{k_{\text{on}}(S_2)} \sum\limits_{n \in S_2}k_{\text{on}}(n) T(n),
\end{equation}
where  $k_{\text{on}}(n)/k_{\text{on}}(S_{2})$ is the probability to bind to DNA segment $S_{2}$ at the site $n$, and  $T(n)$ is the mean first-passage time to reach the target site $m$ starting at the initial position $n$, which  is given by \cite{kolomeisky2013},
\begin{equation}
    T(n)= 
\begin{dcases}
    \sum\limits_{K =n}^{m-1} \left( \sum\limits_{i=l}^{K}\frac{1}{\mu_i}\prod\limits_{j=i+1}^{K}\frac{w_j}{\mu_j} \right) ,& \text{if } n < m \\
    \sum\limits_{K =L+1-n}^{L-m} \left( \sum\limits_{i=l}^{K}\frac{1}{\omega_{L+l-i}}\prod\limits_{j=i+1}^{K}\frac{\mu_{L+l-j}}{w_{L+l-j}}\right),& \text{if } n > m
\end{dcases}
\end{equation}

Under stationary conditions we have $\frac{\partial P_0(t)}{\partial t}=\frac{\partial P_1(t)}{\partial t}=\frac{\partial P_2(t)}{\partial t}=0$, and  Eqs. (7), (8) and (9) can be easily solved, producing stationary probabilities $P_{0}$, $P_{1}$ and $P_{2}$. This allows us to calculate the flux to the target as $J=k_{t} P_2 +k_{\text{on}}(m)P_0$. Finally, the mean search time is given by, 
\begin{equation}
   T=J^{-1}=\left[\frac{1}{k_t}+\frac{k_{\text{on}}(S_1)+k_{\text{off}}(S_1)}{k_{\text{off}}(S_1)k_{\text{on}}(S_2)}\right]\frac{k_{\text{on}}(S_2)}{k_{\text{on}}(S_2)+k_{\text{on}}(m)}. 
\end{equation}

\section{Results}
\label{sec-results}

To test analytical predictions in the limiting cases and to probe the dynamics in the system for general range of parameters, extensive Monte Carlo computer simulations that utilized a Gillespie algorithm  were performed \cite{gillespie1977exact}. We introduce a scanning length, $\lambda_0 \equiv \sqrt {\frac{\mu_0}{k_{\text{off}}^{(0)}}}$, as an important parameter that quantifies the search process. It is defined as the average length the protein slides along the DNA chain before dissociating under no  potential ($G_0(n) = 0$). The actual sliding length will be modified by the effect of the real free-energy profile $G(n)$. The scanning length also correlates with the lifetime of DNA looped configurations: the longer $\lambda_{0}$, the more the systems spends in the looped states.

\begin{figure}[H]
\centering
\includegraphics[width=1.0\columnwidth]{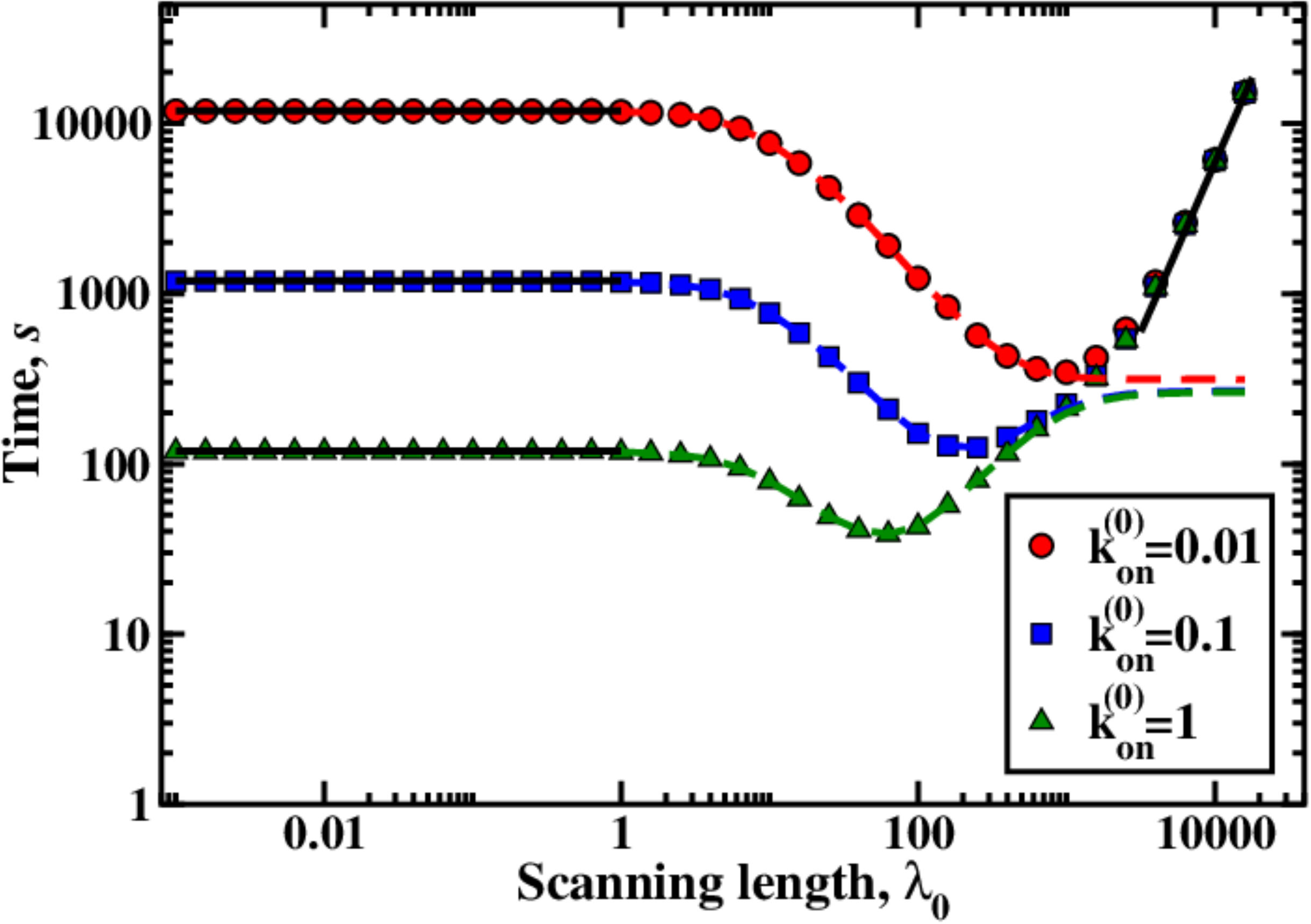}
\caption{Mean search times as a function of the sliding length $\lambda_0$ in the absence (dashed lines) and in the presence of obstacle (symbols and solid curves) for three different values of the protein association rates. The black curves are the analytic results for  2 limiting regimes: for no sliding regime ($\lambda_0 < 1$) and for sliding dominated regime ($\lambda_0 > L$). The following parameters were used in calculations: $\mu_0=60$ s$^{-1}$, $L=m=300$, and $l=150$.}
\label{fig:loopingtime}
\end{figure}

The results of computer simulations and analytical predictions are presented in Fig. \ref{fig:loopingtime} where the mean search times as a function of the scanning length $\lambda_{0}$ are plotted for different association rates. Three search dynamic regimes can be identified. For short scanning lengths ($\lambda_{0} \le 1$), which corresponds to short-lived DNA looped conformations, the search is independent of the obstacle and it also does not depend on the scanning length. This is a ``no-sliding" regime where the protein does not slide much on DNA. In this dynamic regime, our analytical predictions perfectly agree with the results of computer simulations. Increasing the scanning length ($1 \le \lambda_{0} \le L$, intermediate lifetimes for DNA-looped states) accelerates the search dynamics. This is because in addition to coming directly from the solution the protein can also reach the target via sliding along the DNA chain. But the obstacle again does not affect the search dynamics - this can be seen by comparing the results for our model with the obstacle with the model without obstacles that was investigated before (dashed curves in Fig. 4) \cite{shin2019}. The situation dramatically changes in the regime of long scanning lengths ($\lambda_{0} >L$), which corresponds to the long-lived DNA-looped conformations. Here the obstacle significantly slows down the search dynamics. Since in this regime the protein mostly reaches the target via sliding, the protein occasionally maybe trapped in configurations where the obstacle prevents reaching the target (like the region $S_{1}$ in Fig. 3). One could also see that in this dynamic regime our analytical calculations fully agree with computer simulations

The next question we investigated is the role of relative positions of the target and the obstacle in the dynamic regimes with dominating long-lived DNA loops. The results are presented in Fig. 5 where the target position is varied while the obstacle position is fixed in the middle of the DNA stand. The search time is significantly larger if the target is located for $1 \le m < l$ in comparison with positioning the target to $l < m \le L$: in Fig. 4 we employed $l=150$ and $L=300$. To understand this result, let us recall that according to Fig. 2 it costs much more energy to make short loops ($1 \le n \le 150$), and this means that the protein has a lower probability to reach this part of DNA. The protein can also dissociate more easily from this region. So putting the target in this segment will slow down the search due to slow association rates and multiple visits. But the segment $150 \le n \le 300$ is energetically much more favorable, and putting  the target there will lead to faster search due to faster association rates and not so many repeat visits. Our calculations also show that in all situations for large scanning lengths (long-lived DNA loops) the search time in the presence of the obstacle is always slower than in the case without of the obstacle, as illustrated in Fig. 5.

\begin{figure}[H]
\centering
\includegraphics[width=1.0\columnwidth]{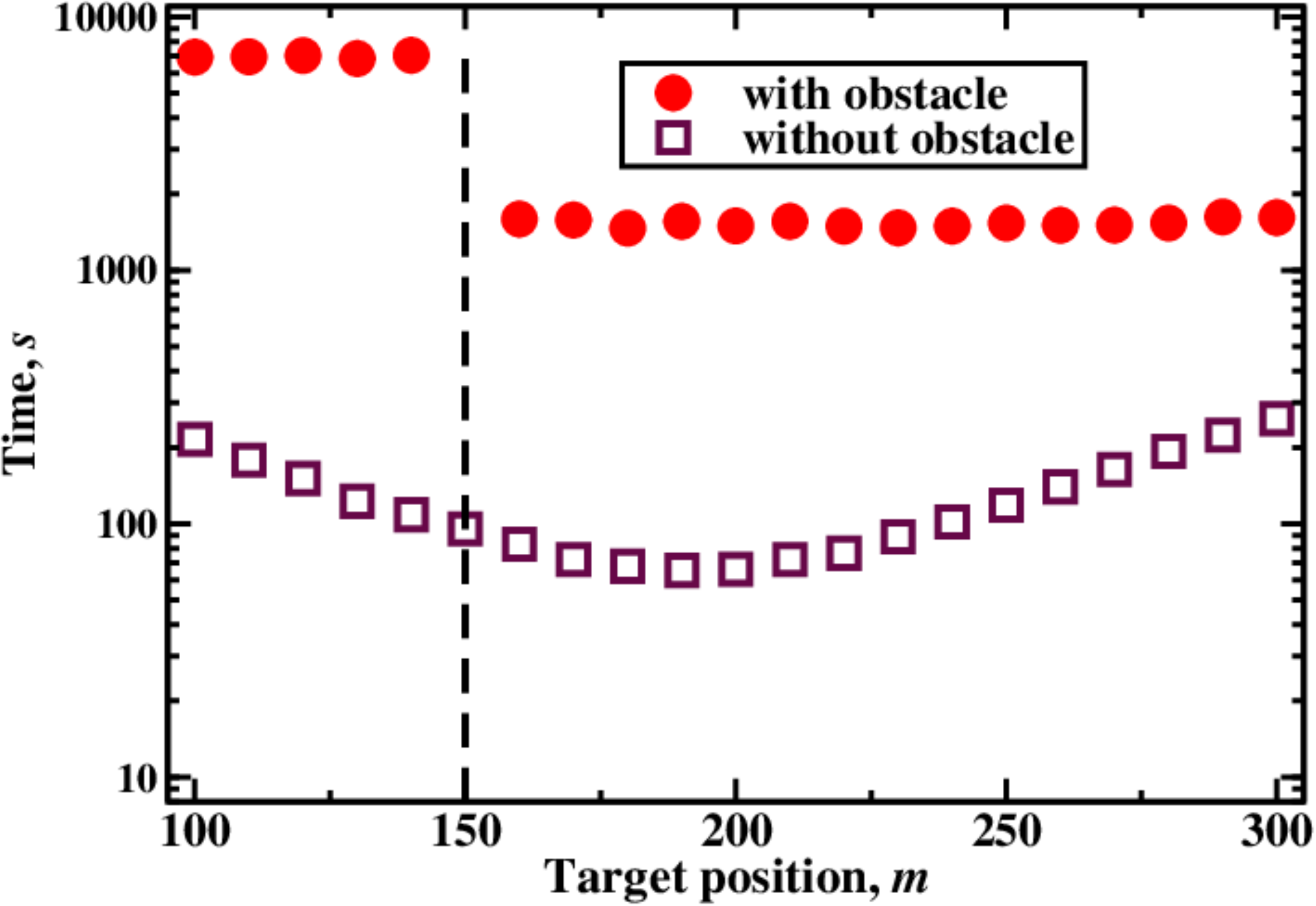}
\caption{Mean search time as a function of the target position $m$ for large scanning length with an obstacle (filled circles) and without an obstacle (empty squares). The following parameters were used in calculations: $\lambda_{0}=5000$, $l=150$, $k_{\text{on}}^{(0)}=1$ s$^{-1}$, $L=300$ and $\mu_0=60$ s$^{-1}$. The vertical dashed line indicates the obstacle position.}
\label{fig:4}
\end{figure}

In the related study, we fixed the position of the target $m$ and varied the position of the obstacle $l$, and the results are illustrated in Fig. 6 where the target is at the DNA end, $m=L$. When the obstacle is found far away from the target at the other end of DNA, the search is not affected by the presence of the obstacle. This is because the formation of short loops is energetically unfavorable so that the protein does not go frequently to this region. In addition, the obstacle is too far away from the target to affect the sliding in the direction of the target. However, moving the obstacle closer to the target increases the mean search time since the obstacle now works as the barrier that cannot be passed over. The protein must dissociate more frequently in order not to be trapped in the region $n<l$.

\begin{figure}[H]
\centering
\includegraphics[width=1.0\columnwidth]{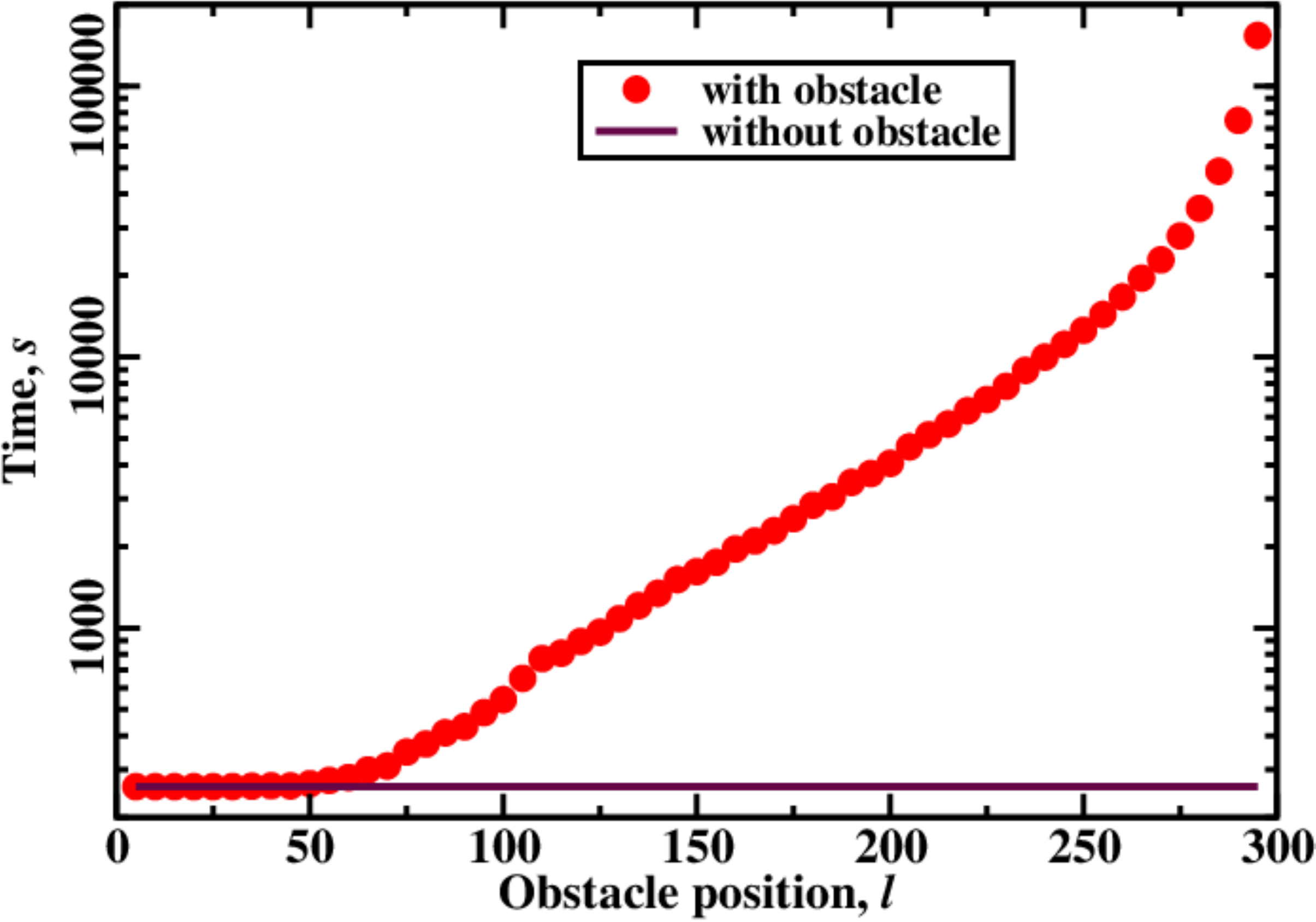}
\caption{Mean search time as a function of the obstacle position $l$ for the fixed position of the target. Circles are for the model with the obstacle, while the solid line is for the model without obstacle for the same conditions. The following parameters were used in calculations: $\lambda_{0}=5000$,  $L=300$, $m=300$, $k_{\text{on}}^{(0)}=1$ s$^{-1}$ and $\mu_0=60$ s$^{-1}$.}
\label{fig:5}
\end{figure}

\begin{figure}[H]
\centering
\includegraphics[width=1.0\columnwidth]{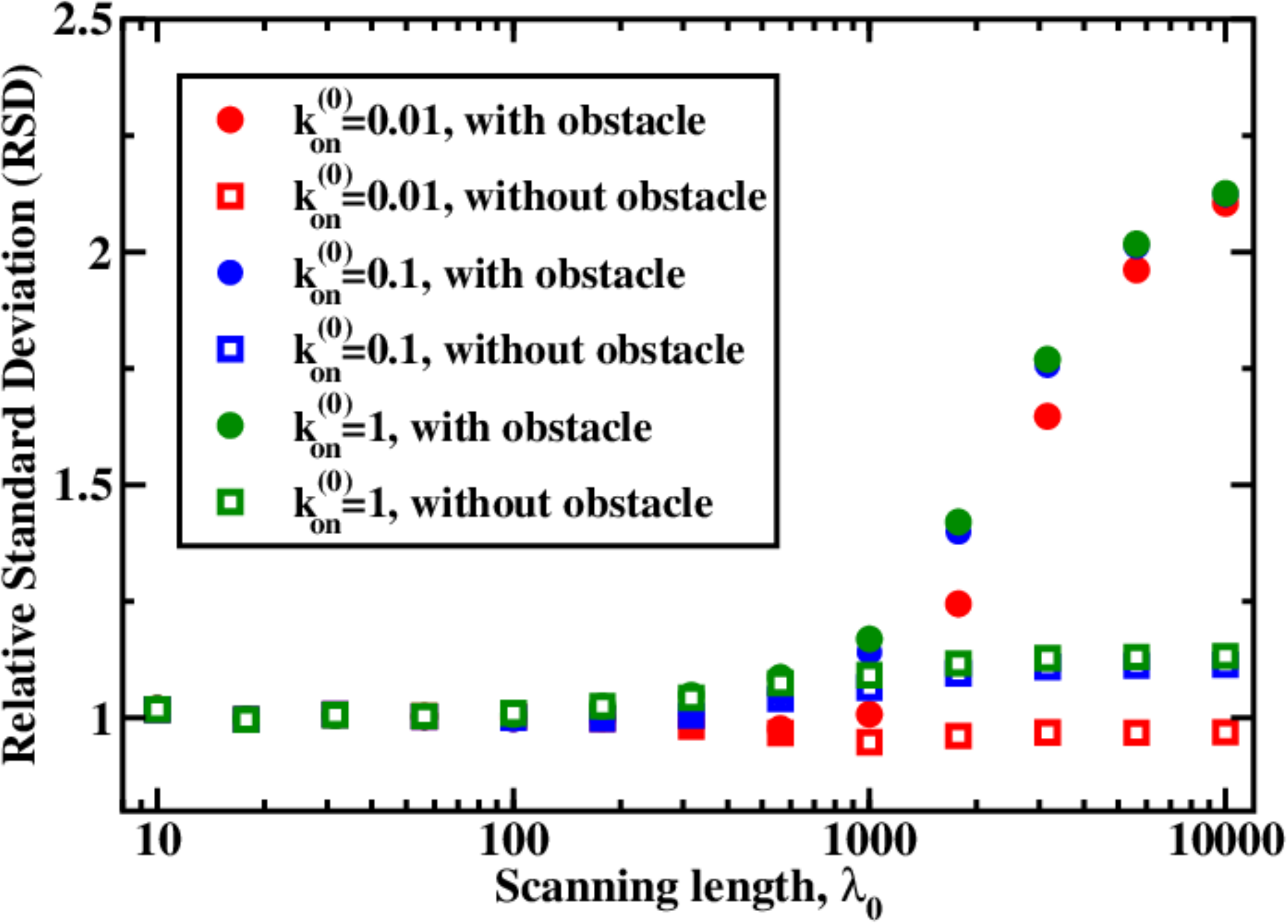}
\caption{Relative standard deviation  of the search times as a function of the scanning length $\lambda_0$ for different values of association rates. Filled circles correspond to the systems with the obstacle, empty squares describe the systems without obstacle. The following parameters were used in calculations: $L=300$, $m=300$, $l=150$ and $\mu_0=60$ s$^{-1}$.}
\label{fig:6}
\end{figure}

Until now, we focused on the mean times as a measure of the searching dynamics of multi-site proteins, but the distribution of search times should provide a more comprehensive description of the process  \cite{pulkkinen2013distance,godec2016universal,grebenkov2018strong}. In order to understand how the presence of the obstacle influences the distribution of search times, we analyzed the relative standard deviation (RSD), which is the ratio of the standard deviation of search times to the mean search time, and the results are presented  in Fig. \ref{fig:6}. This quantity also measures the degree of stochastic noise in the system. One can see that for short and intermediate scanning lengths, which correspond to the case when the lifetimes of the DNA-looped states is not large, the RSD is close to 1, and it does not depend on the presence of the obstacle. We also find that the distribution of search times is close to the exponential with relatively small degree of noise in the system. However, for the regime when the long-lived DNA loops dominate (large scanning lengths) the presence of the obstacle affects the distribution of search times. The distribution becomes much wider, which also corresponds to the increase in the stochastic noise in the system. This result is easy to understand because in this regime the protein will frequently visit the DNA segment from which it cannot slide to the target, and multiple visits will be done until the goal is accomplished. This should increase the noise in the system. We also notice that RSD is almost independent of the association rates.

Our theoretical calculations provide important molecular insights on the role of obstacles in the multi-site protein target search. They suggest that from the biological point of view the presence of obstacle does not always lead to significant delays in the formation of protein-DNA complexes, in contrast to naive expectations. If the DNA loops lifetimes are not too short and too long, the system can provide a fast target search and it can also avoid the negative effect of the obstacle. This range of parameters also do not increase the noise in the system which is important for robustness of biological systems. In addition, the negative effects of the obstacles can be moderated by proper arrangements of specific target sites with respect to them. It will be interesting to check if real biological systems satisfy these conditions.

\section{Summary and Conclusions}
\label{sec-conclusion}

In this work, a theoretical analysis on the role of static obstacles in the formation of protein-DNA complexes with multiple interaction sites is presented. We concentrated on the final stages of the multi-site protein target search that involves the formation of DNA loops and sliding in the looped configurations. A discrete-state stochastic model for this process is developed and it is investigated using exact analytical calculations and Monte Carlo computer simulations.  It is found that the static obstacle does not interfere with the target search dynamics if the lifetimes of the DNA-looped configurations are not too long. However, in the situations when the DNA loops are long lived the search dynamics is significantly slowed down, and the stochastic noise in the system also increases. In addition, the search dynamics might be affected by the relative location of the target and the obstacle. It is argued that biological system might prefer the conditions with intermediate lifetimes of DNA-looped states since it leads to the fastest search dynamics without much effect from the obstacles. 

Although our theoretical approach identifies the most relevant features of the multi-site protein search dynamics in the presence of obstacles, we note that the presented theoretical picture is rather limited with many approximations. Our method does not take into account the sequence heterogeneity of DNA molecules, while this might be an important effect \cite{shvets2015}. In addition, more advanced descriptions of the free-energy profile for polymers with loops is available. Our theoretical analysis also will not work for very long DNA chains because the DNA chain will not be able to relax fast enough. Another complication is the assumption of no correlation in the protein consecutive bindings. Furthermore, our theoretical approach neglects the possibility of DNA supercoiling and twists, which should complicate the overall search dynamics. Despite these considerations, the presented theoretical model clarifies some important molecular details of very complex biological processes of formation of protein-DNA complexes. It will be crucial to test our theoretical conclusions in experiments as well as in more advanced theoretical studies.

\section*{Acknowledgments}

The work was supported by the Welch Foundation (Grant C-1559), by the NSF (Grant CHE-1664218), and by the Center for Theoretical Biological Physics sponsored by the NSF (Grant PHY-1427654).

%%%REFERENCES%%%
\bibliography{mybib}

\begin{thebibliography}{10}

\bibitem{alberts2014molecular}
B.~Alberts {\em et~al.},
\newblock NY: WW Norton \& Company  (2014).

\bibitem{lodish2008molecular}
H.~Lodish {\em et~al.},
\newblock {\em Molecular cell biology} (Macmillan, 2008).

\bibitem{grindley2006mechanisms}
N.~D. Grindley, K.~L. Whiteson, and P.~A. Rice,
\newblock Annu. Rev. Biochem. {\bf 75}, 567 (2006).

\bibitem{halford2004}
S.~E. Halford and J.~F. Marko,
\newblock Nucleic Acids Research {\bf 32}, 3040 (2004).

\bibitem{mani2010triggers}
R.-S. Mani and A.~M. Chinnaiyan,
\newblock Nature Reviews Genetics {\bf 11}, 819 (2010).

\bibitem{bushman2005genome}
F.~Bushman {\em et~al.},
\newblock Nature Reviews Microbiology {\bf 3}, 848 (2005).

\bibitem{cournac2013dna}
A.~Cournac and J.~Plumbridge,
\newblock Journal of bacteriology {\bf 195}, 1109 (2013).

\bibitem{riggs1970lac}
A.~D. Riggs, S.~Bourgeois, and M.~Cohn,
\newblock Journal of Molecular Biology {\bf 53}, 401 (1970).

\bibitem{berg1976association}
O.~G. Berg and C.~Blomberg,
\newblock Biophysical Chemistry {\bf 4}, 367 (1976).

\bibitem{winter1981diffusion}
R.~B. Winter and P.~H. Von~Hippel,
\newblock Biochemistry {\bf 20}, 6948 (1981).

\bibitem{von1989facilitated}
P.~H. von Hippel and O.~G. Berg,
\newblock Journal of Biological Chemistry {\bf 264}, 675 (1989).

\bibitem{coppey2004}
M.~Coppey, O.~B{\'e}nichou, R.~Voituriez, and M.~Moreau,
\newblock Biophysical Journal {\bf 87}, 1640 (2004).

\bibitem{lomholt2009}
M.~A. Lomholt, B.~van~den Broek, S.-M.~J. Kalisch, G.~J. Wuite, and R.~Metzler,
\newblock Proceedings of the National Academy of Sciences USA {\bf 106}, 8204
  (2009).

\bibitem{kolomeisky2013b}
A.~Veksler and A.~B. Kolomeisky,
\newblock The Journal of Physical Chemistry B {\bf 117}, 12695 (2013).

\bibitem{esadze2014positive}
A.~Esadze, C.~A. Kemme, A.~B. Kolomeisky, and J.~Iwahara,
\newblock Nucleic acids research {\bf 42}, 7039 (2014).

\bibitem{mirny2009}
L.~Mirny {\em et~al.},
\newblock Journal of Physics A: Mathematical and Theoretical {\bf 42}, 434013
  (2009).

\bibitem{benichou2011}
O.~B{\'e}nichou, C.~Loverdo, M.~Moreau, and R.~Voituriez,
\newblock Reviews of Modern Physics {\bf 83}, 81 (2011).

\bibitem{kolomeisky2011}
A.~B. Kolomeisky,
\newblock Physical Chemistry Chemical Physics {\bf 13}, 2088 (2011).

\bibitem{sheinman2012}
M.~Sheinman, O.~B{\'e}nichou, Y.~Kafri, and R.~Voituriez,
\newblock Reports on Progress in Physics {\bf 75}, 026601 (2012).

\bibitem{kolomeisky2016}
A.~A. Shvets and A.~B. Kolomeisky,
\newblock The Journal of Physical Chemistry Letters {\bf 7}, 5022 (2016).

\bibitem{shin2019}
J.~Shin and A.~Kolomeisky,
\newblock Soft matter {\bf 15}, 5255 (2019).

\bibitem{marklund2013transcription}
E.~G. Marklund {\em et~al.},
\newblock Proceedings of the National Academy of Sciences {\bf 110}, 19796
  (2013).

\bibitem{brackley2013}
C.~Brackley, M.~Cates, and D.~Marenduzzo,
\newblock Physical Review Letters {\bf 111}, 108101 (2013).

\bibitem{shvets2015role}
A.~Shvets, M.~Kochugaeva, and A.~B. Kolomeisky,
\newblock The Journal of Physical Chemistry B {\bf 120}, 5802 (2015).

\bibitem{shvets2016crowding}
A.~A. Shvets and A.~B. Kolomeisky,
\newblock The journal of physical chemistry letters {\bf 7}, 2502 (2016).

\bibitem{gomez2016facilitated}
D.~Gomez and S.~Klumpp,
\newblock Physical Chemistry Chemical Physics {\bf 18}, 11184 (2016).

\bibitem{tafvizi2008tumor}
A.~Tafvizi {\em et~al.},
\newblock Biophysical journal {\bf 95}, L01 (2008).

\bibitem{shvets2016role}
A.~A. Shvets and A.~B. Kolomeisky,
\newblock The journal of physical chemistry letters {\bf 7}, 5022 (2016).

\bibitem{shin2018surface}
J.~Shin and A.~B. Kolomeisky,
\newblock The Journal of Physical Chemistry B {\bf 122}, 2243 (2018).

\bibitem{kolomeisky2013}
X.~Li and A.~B. Kolomeisky,
\newblock The Journal of Chemical Physics {\bf 139}, 144106 (2013).

\bibitem{gillespie1977exact}
D.~T. Gillespie,
\newblock The Journal of Physical Chemistry {\bf 81}, 2340 (1977).

\bibitem{pulkkinen2013distance}
O.~Pulkkinen and R.~Metzler,
\newblock Physical review letters {\bf 110}, 198101 (2013).

\bibitem{godec2016universal}
A.~Godec and R.~Metzler,
\newblock Physical Review X {\bf 6}, 041037 (2016).

\bibitem{grebenkov2018strong}
D.~S. Grebenkov, R.~Metzler, and G.~Oshanin,
\newblock Communications Chemistry {\bf 1}, 96 (2018).

\bibitem{shvets2015}
A.~A. Shvets and A.~B. Kolomeisky,
\newblock The Journal of Chemical Physics {\bf 143}, 245101 (2015).

\end{thebibliography}
\bibliographystyle{h-physrev} %the RSC's .bst file

\end{document}